\documentclass[MSNbibl,nameyear,dvips]{arxstspdf}
\usepackage{flushend}
\usepackage{stfloats}
\usepackage{graphicx}

\volume{26}
\issue{4}
\pubyear{2011}
\firstpage{601}
\lastpage{603}
\doi{10.1214/11-STS356B}
\referstodoi{10.1214/11-STS356}

\begin{document}
\begin{frontmatter}
\vspace*{6pt}
\title{Discussion of ``Multiple Testing for Exploratory Research'' by J. J. Goeman and A. Solari}
\runtitle{Discussion}

\begin{aug}
\author[a]{\fnms{Nicolai} \snm{Meinshausen}\corref{}\ead[label=e1]{meinshausen@stats.ox.ac.uk}}
\runauthor{N. Meinshausen}

\affiliation{University of Oxford}

\address[a]{Nicolai Meinshausen is University Lecturer, Department of Statistics,
University of Oxford, UK \printead{e1}.}

\end{aug}



\end{frontmatter}
I want to congratulate the authors on this thought-provoking and
important paper on multiple testing in exploratory settings.

Standard Multiple Testing procedures can appear very mechanistic.
Hypotheses are ordered by increasing $p$-value. Given a Type I error
criterion, the Multiple Testing procedure selects a cut-off in this list.
Simply working down the list of hypotheses in order of their $p$-values
is perhaps suboptimal for exploratory analysis as a lot of information
is lost in this way and important discoveries might be missed. Some
previous work has addressed this issue by\break changing the ranking of the
hypotheses. To highlight only three examples: Tibshirani and Wasserman
(\citeyear{tibshirani2006correlation}) devised a method to borrow strength across
highly correlated test statistics in microarray experiments.
Storey (\citeyear{storey2007optimal}) proposed an ``optimal discovery'' procedure that
again leads to a different ranking of variables than the ranking
implied by the marginal $p$-values. One of the authors also proposed a
very powerful way of incorporating known network structure into the
testing procedure [Goeman and Mansmann, \citeyear{goeman2008multiple}].

The proposed approach to exploratory multiple testing is more radical,
though, than changing the cut-off or
changing the ranking of hypotheses. Instead of the perhaps rather dull
task of selecting a~cut-off in a list of ordered hypotheses, the
researcher can reject for follow-up analysis any set of hypotheses he
or she regards as interesting, using all the information at hand. The
method then returns a~lower bound on the number of false null\vadjust{\goodbreak}
hypotheses (true discoveries) in this set. Since the bound is valid
simultaneously across all sets, an exploratory approach does not
invalidate the error bound.

I think this method will be very important and useful in many fields as
it allows a flexible exploration of possibly interesting sets of
hypotheses, while at the same time protecting the practitioner against
too many false rejections
(or at least managing expectations about the number of true discoveries
one can hope to make).

\begin{figure*}

\includegraphics{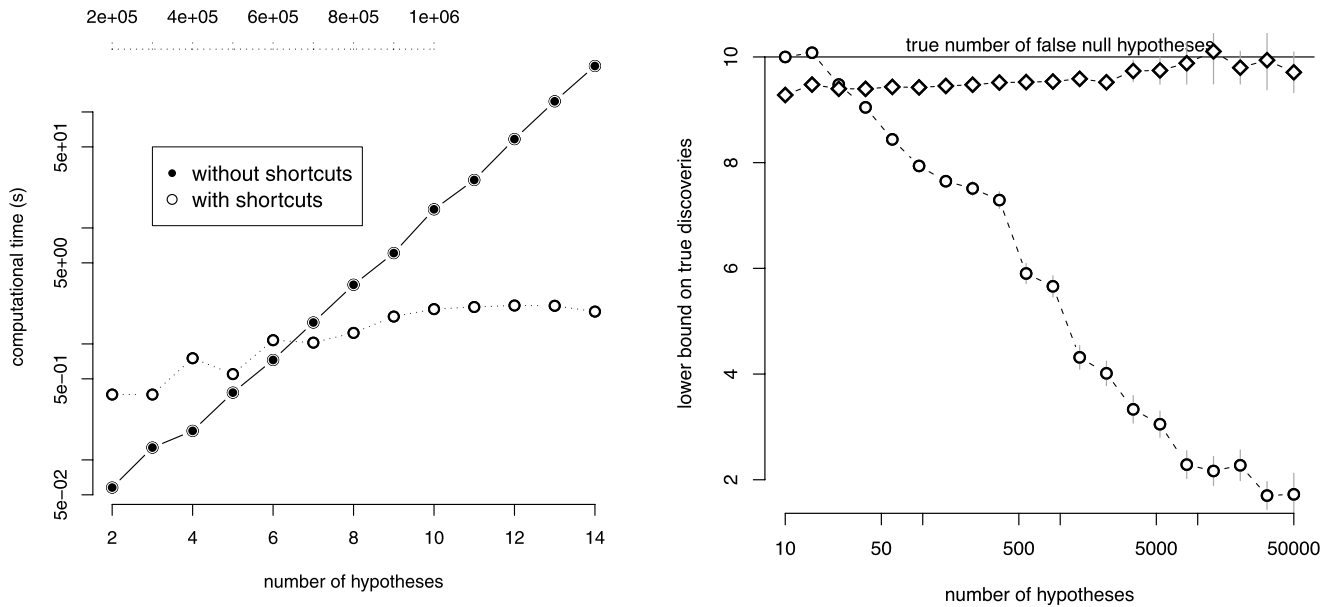}

\caption{Left: computational time (on log-scale) for the
standard approach in regression (solid line), with the number of
hypotheses ranging from 2 to 12, and computational time for the
shortcut for independent tests (dotted line), with the number of
hypotheses between $2\cdot10^5$ and $1.2\cdot10^6$. Only the latter
is feasible for large-scale testing. Right: the average lower bound on
the total number of true discoveries with the proposed approach
(circles) and an alternative given in the text (diamonds). Due to the
simultaneous nature of the proposed bounds, the overall power to reject
is good for a few dozen hypotheses but is deteriorating as the number
of hypotheses grows into the hundreds.}\label{fig:1}
\end{figure*}

There is a price to be paid for the simultaneous nature of the bound,
though. I have some doubts (hopefully unfounded) about the
applicability to\break large-scale testing situations as they arise, for
example, in genomics or astronomy for two reasons: computational
complexity and statistical power.

It is obvious and also acknowledged by the authors that the proposed
procedure without shortcuts will be impractical for even just a few
dozen hypotheses. The computational complexity is simply too high. An
example is shown in Figure~\ref{fig:1} for a genomics regression
example with less than one hundred observations. The proposed method
takes already more than half a minute for 12 predictor variables on
a~standard computer with a 3~GHz CPU and the supplied \texttt{cherry}
R-package and the complexity seems to be (super-)exponential in the
number of hypotheses, as one would expect.
The proposed shortcuts are not applicable in all settings. If they are
applicable, they seem to be very effective in reducing the
computational complexity, making large-scale testing feasible.
Figure~\ref{fig:1} shows that even testing situations with $>10^6$
tests are handled in about a second or less.

Maybe more worrying, the statistical power of the method deteriorates
with an increasing number of hypotheses. This is due to the
simultaneous nature of the bound on the number of correctly rejected
hypotheses among all possible sets of hypotheses. I~compared the power
for a simple setting, in which there are $m$ independent $p$-values
$p_i$ with $i=1,\ldots,m$ with distribution $p_i \sim U([0,c_i])$ and
$c_i=1$ if $i>10$ and $c_i=0.1/ m$ if $i\le10$ (there are hence 10
false null hypotheses). If rejecting all hypotheses, the lower bound
for the number of correctly rejected hypotheses is shown as a function
of $m$ in Figure \ref{fig:1}, along with the bound for the same
quantity proposed by
Meinshausen and Rice (\citeyear{meinshausen04estimating}). The proposed approach works very well
up to a few dozen hypotheses. If the number of hypotheses is in the
hundreds, the number of sets the bound needs to be valid over is
getting so large that the power of the method starts to deteriorate quickly.

I acknowledge that the comparison is not quite fair since the method in
Meinshausen and Rice (\citeyear{meinshausen04estimating}) does much less: it only gives a lower
bound on the \emph{total} number of false null hypotheses or a lower
bound for the number of true discoveries in a list that is ordered by
increasing $p$-values of the hypotheses. [If we were to ask only if
there are any false null hypotheses at all, we could be even more
sensitive to deviations from the global null hypothesis with Higher
Criticism (Donoho and Jin, \citeyear{donoho04higher}).]
And for fewer than 50 hypotheses, the proposed bound is remarkably good.

The power and computational cost objectives thus both indicate that the
method is working very well for up to a few dozen hypotheses but will
probably need refinements for large-scale testing.

A thought regarding the presentation of results. As proposed, the
method acts somewhat like a black-box: if given a set of hypotheses, it
returns a lower bound on the number of true discoveries within this set.
While this might be the right approach in many exploratory settings, I
also think that many practitioners could use some guidance as to which
sets of hypotheses could be interesting (without prescribing exactly
which ones to reject, so as to not fall back into the standard ranking scheme).
A step in this direction is the helpful concept of \emph{defining
hypotheses}, which summarizes the results of the procedure in compact form.

Each defining hypothesis is a set of hypotheses out of which at least
\emph{one hypothesis} must be a false null hypothesis. In other words:
the defining hypotheses have a logical AND--OR connection (with AND
between the sets of hypotheses and OR between hypotheses in a set). A
complementary view could be given by a logical OR--AND connection, with
OR between sets of hypotheses and AND between hypotheses in a set. The
results are still presented as sets of hypotheses. Among all these sets
and conditional on event $E$, there is now guaranteed to be at least
\emph{one set} such that \emph{all hypotheses} in this set are false.
This extends the usual Multiple Testing paradigm, where the user is
handed back just one set of hypotheses, which is guaranteed to be a set
of false null hypotheses.

\begin{figure}

\includegraphics{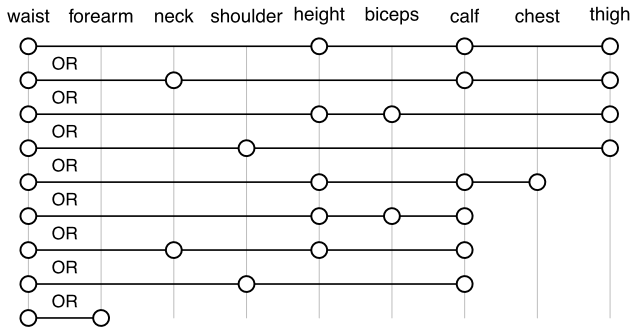}

\caption{An alternative presentation of the results in
the regression example. All dots on each horizontal line form a set of
hypotheses. Out of all nine sets, at least one must correspond to a set
of false null hypotheses (conditional on event~$E$).}\label{fig:2}
\end{figure}

In the regression example, there are nine such sets, the first two
being \{\textit{waist, height, calf, thigh}\} and \{\textit{waist,
neck, calf, thigh}\}.
Figure~\ref{fig:2} visualizes them. Among these nine sets, at least one
must be a set where all hypotheses are false null hypotheses (always
conditional on the event $E$). We can then directly read off that if
\textit{height} is known to be a~null hypothesis (either by a follow-up
experiment or through prior knowledge), then the results give no reason
any longer to suppose that \textit{chest} was a~false null (since
\textit{chest} is only part of the fifth set; and if \textit{height} is
a true null, this set can be excluded and neck will not any longer be
in the union of all other candidate sets). Or, if \textit{calf} can be
excluded, then the results do not give reason to still suspect that
\textit{neck} was a false null hypothesis. Such statements and
connections are much more difficult to read off the set of defining
hypotheses but might be useful in practice, when planning which
hypotheses to follow up.

I want to congratulate the authors again on this very impressive and
useful paper and I hope to see strong uptake of the method.\vspace*{-3pt}

\section*{Acknowledgment}
The author wishes to acknowledge support from the Leverhulme Trust.\vspace*{15pt}

%

\end{document}